\documentclass[sigconf]{acmart}

\usepackage{color}
\usepackage{hyperref}

\usepackage{booktabs} 

\setcopyright{rightsretained}

\acmDOI{10.475/123_4}

\acmISBN{123-4567-24-567/08/06}

\acmConference[RecSys '17]{ACM Recommender Systems conference}{27th-31st August 2017}{Como, Italy} 	
\acmYear{2017}
\copyrightyear{2017}

\acmPrice{15.00}

\begin{document}

\title{Contextual Sequence Modeling for Recommendation with Recurrent Neural Networks}

\author{Elena Smirnova}
\affiliation{%
  \institution{Criteo Research}  
  \city{Paris}   
}
\email{e.smirnova@criteo.com}

\author{Flavian Vasile}
\affiliation{%
  \institution{Criteo Research}  
  \city{Paris}   
}
\email{f.vasile@criteo.com}

\begin{abstract}
Recommendations can greatly benefit from good representations of the user state at recommendation time. Recent approaches that leverage Recurrent Neural Networks (RNNs) for session-based recommendations have shown that Deep Learning models can provide useful user representations for recommendation. However, current RNN modeling approaches summarize the user state by only taking into account the sequence of items that the user has interacted with in the past, without taking into account other essential types of context information such as the associated types of user-item interactions, the time gaps between events and the time of day for each interaction. To address this, we propose a new class of Contextual Recurrent Neural Networks for Recommendation (CRNNs) that can take into account the contextual information both in the input and output layers and modifying the behavior of the RNN by combining the context embedding with the item embedding and more explicitly, in the model dynamics, by parametrizing the hidden unit transitions as a function of context information.  We compare our CRNNs approach with RNNs and non-sequential baselines and show good improvements on the next event prediction task.
\end{abstract}



\maketitle

\section{Introduction}
\label{sec:intro}

Current recommendation approaches can be broadly categorized based on their prediction task into  missing link prediction methods (what should the user see that he/she has not seen yet?) and next item prediction methods (what are the most likely continuations of the user activity?). 

The first type of methods includes factor models that learn a low-rank decomposition of sparse user-item interactions matrix \cite{Koren:2009:MFT:1608565.1608614, prod2vec, metaprod2vec}. The missing entries are then extrapolated as the inner product between the resulting user and item latent vectors.

In this paper, we are interested in next item prediction methods. Typically, those methods look at the organic user behavior and build a predictor of the next event based on the last K user actions \cite{Sarwar:2001:ICF:371920.372071}.

The simplest way to deal with sequential prediction is to modify the non-sequential models to include sequence information. In particular, 2015 RecSys Challenge winners \cite{romov2015recsys} augmented a gradient boosting algorithm with time features and counters. 

Another popular approach, called neighborhood methods, is to consider only the last item in the user sequence to produce the recommendation \cite{Sarwar:2001:ICF:371920.372071}. These methods compute co-occurrences of items within some scope, typically, user session. However, an attempt to go beyond the history of size one with neighborhood methods results in a sparsity problem as co-occurrence with sub-sequences becomes rare.

Instead of relying on features or co-occurrence frequencies, sequence methods directly model the sequence of user actions. In particular, Recurrent Neural Networks (RNNs), a class of deep neural networks, have been successfully applied to the task of next item prediction \cite{rnn:balazs}. RNNs feature hidden state with nonlinear dynamics that enable them to discover patterns of activity that are predictive of next item id.   

Besides the order of items, additional information about user-item interaction is frequently available, such as type of the interaction, the time gaps between events and time of day of interaction. 

\begin{figure}
\begin{center}
  \includegraphics[scale=0.4]{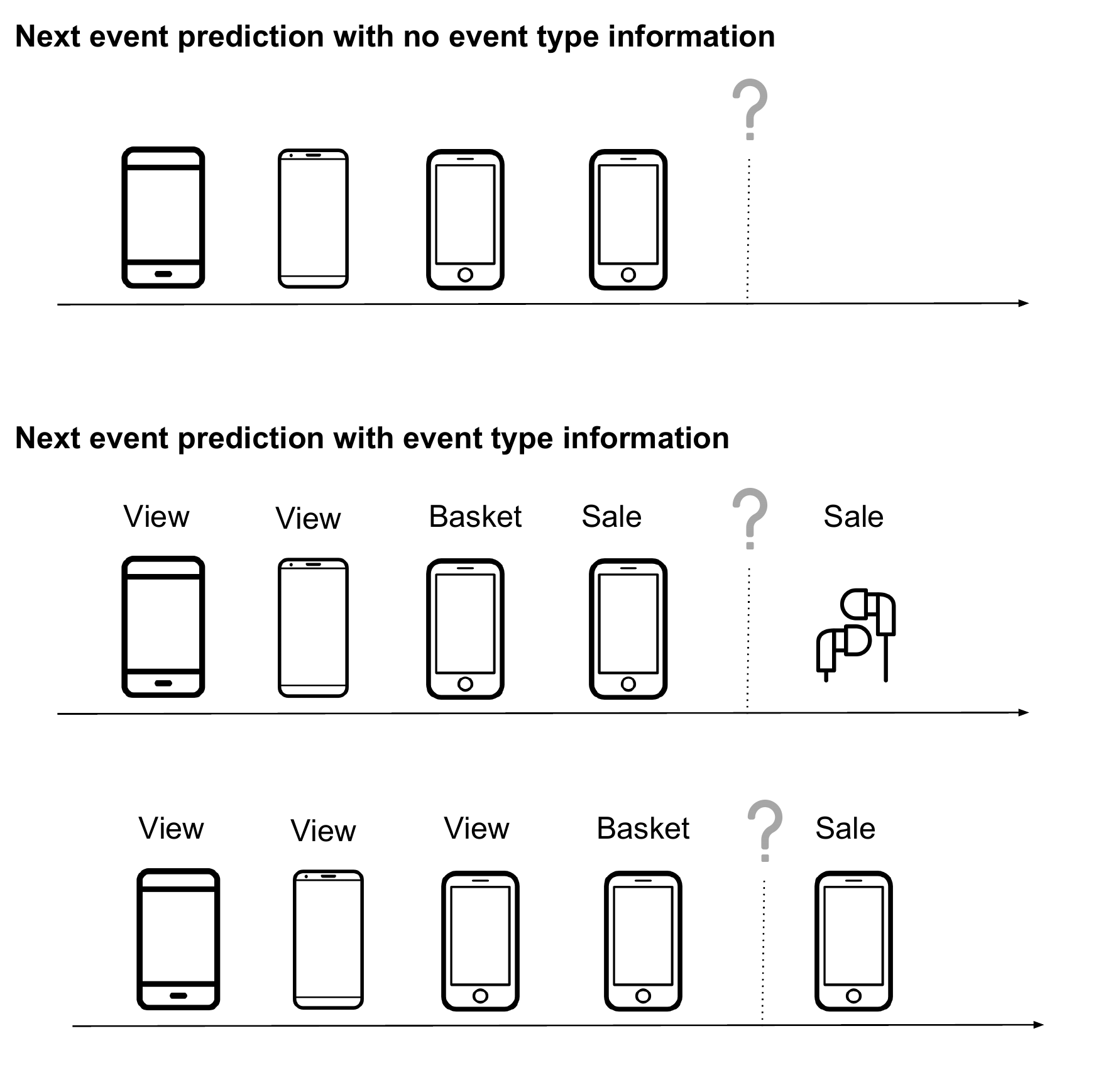}
  \caption{A case where the difference in event types over the same sequence of past item ids leads to big differences in the most likely next item. The top unlabeled sequence represents the information that a standard RNN would have available, leading to an average prediction between the two possible outcomes below. In the labeled sequence below, we observe that the user bought the phone in the last event, so the most likely next item to visit is a complementary item, such as headphones. In the bottom labeled sequence, the used added the phone to basket so the next most likely event is for the user to buy the phone.}
  \label{fig:example_event_type}
 \end{center}
\end{figure}

\begin{figure}
\begin{center}
  \includegraphics[scale=0.4]{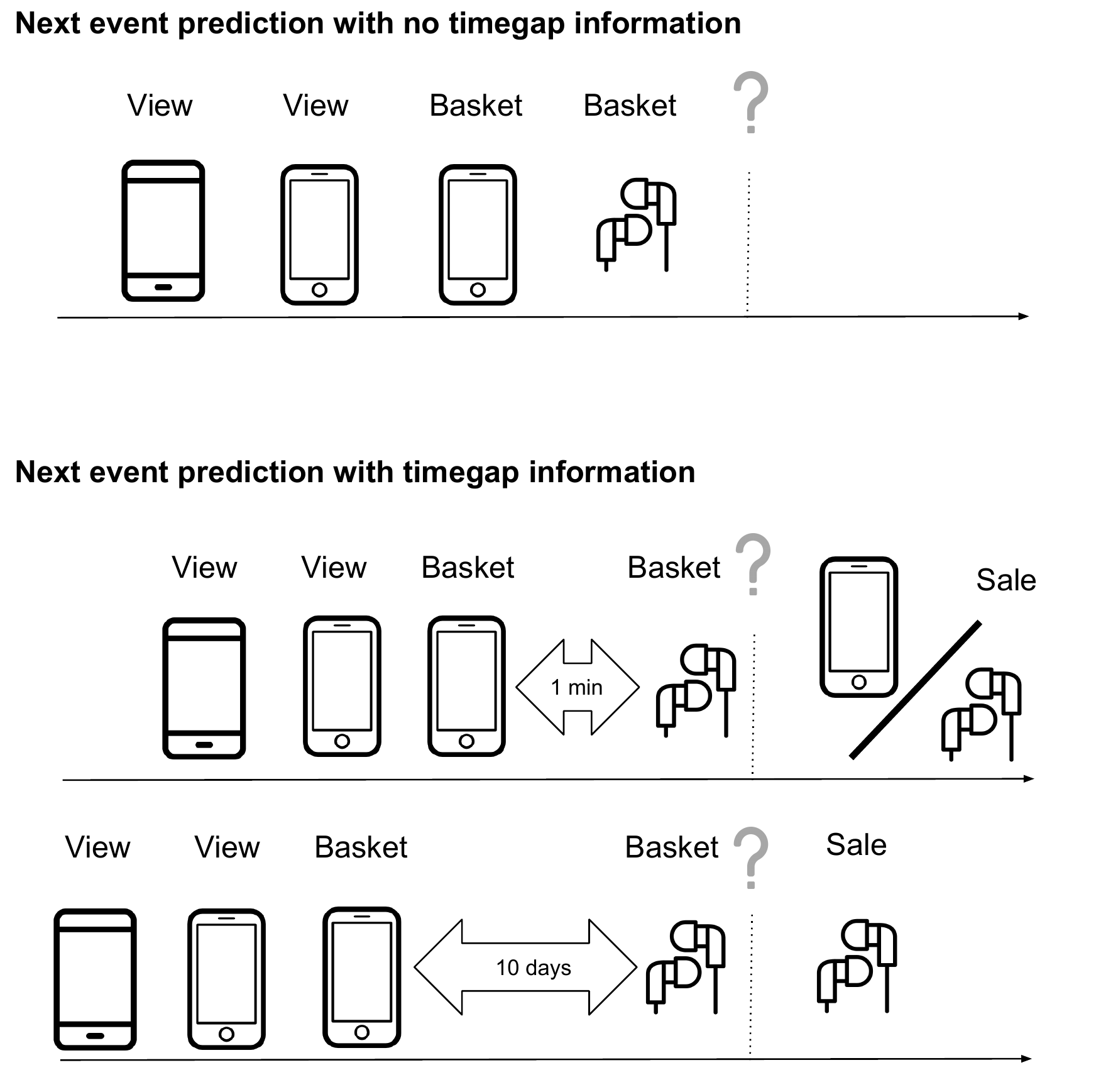}
  \caption{A case where the difference in time gaps over the same sequence of past item ids leads to big differences in the most likely next item. This time knowing that the last event is separated from the other previous events by a large gap, leads to a large change in likelihood.}
  \label{fig:example_time_gap}
 \end{center}
\end{figure}

This contextual information has the potential to greatly improve the next event prediction. For example, knowing the event type of past products can increase or decrease the probability of the identity of the next product the user will interact with. For example, if the user just added a product to basket, then he/she is very likely to buy it next. On the other hand, if the last event is a sale, then it is likely that the next event will be a complementary product rather than the same product. This example is illustrated in Figure \ref{fig:example_event_type}. Similarly, we can observe big differences in the likelihood of the next product given different patterns of time gaps between past user events, shown in Figure \ref{fig:example_time_gap}.

Because we strongly believe context matters in sequence prediction we formally introduce the problem of \emph{Contextual Sequential Modeling for Recommendation} that generalizes the existing setups of Context-Aware Recommendation (CAR) and Sequential Modeling. 

As a solution, we introduce the family of \emph{Contextual RNNs (CRNNs)} algorithms, that can leverage context information for the item sequence modeling in order to improve next item prediction. We study two ways of introducing context in sequence models one of \emph{context-dependent input/output modeling}, where item representation is combined with context through a series of non-linear transformations, and one of \emph{context-dependent dynamics modeling}, where context is used to parametrize the dynamics of the hidden state transition.

We compare our architectures with sequential and non-sequential baselines and show that our method significantly outperforms all of the other methods on two e-commerce datasets, the YooChoose open dataset and an internal dataset.

To summarize, our main contributions are:
\begin{itemize}
	\item The introduction of CRNNs for the problem of contextual sequence modeling for recommendation,
	\item The comparative study of two possible families of CRNN architectures and their benchmarking against previous methods on two real-world e-commerce datasets.
\end{itemize}
 
In Section \ref{sec:relatedwork} we cover previous related work and the relationship with our method. In Section \ref{sec:proposedapproach} we present the CRNN model. In Section \ref{sec:experiments} we present the experimental setup and the results. In Section \ref{sec:conclusion} we summarize our findings and conclude with future directions of research.

\section{Related Work}
\label{sec:relatedwork}

\subsection{Recurrent Neural Networks for Recommendation}
\label{sec:rnn-for-reco}
Hidasi et al. \cite{rnn:balazs} applied RNNs to the task of session-based recommendation. They trained a Gated Recurrent Unit (GRU) \cite{gru:cho} architecture with ranking loss and evaluated on predicting the next item in the user session on e-commerce website. Authors showed that RNNs significantly outperform the CF baseline. 

In a follow-up paper \cite{feature-rnn:balazs}, Hidasi et al. introduced 3 architectures to model multi-modal product representations in sessions. Specifically, in addition to product identifier they modeled product text and image information. The best architecture allowed for interaction of hidden states of different modalities.

Neural architecture proposed in \cite{Twardowski:2016:MCI:2959100.2959162} produced an embedding of sequence of events using RNN to predict the next item represented as a set of attributes. The contextual information was concatenated with associated item information. In experiments, the model with events context outperformed the model based only on items.

\subsection{Context-aware Recommendation}
\label{sec:car}
Numerous studies suggest that context influences user preferences at recommendation. For example, shoppers change their behavior during weekdays and weekends \cite{yoochoose:zalando}. 
Context also changes the item similarities. For example, in the context of a sale, items similar to a sold item are complementary items \cite{Hidasi:2012:FAT:2405742.2405749}. 

Factorization is a primary method of introducing context into recommendation systems. Hidasi and Tikk \cite{car:balazs} introduced a general framework that models interactions between user, items and context in the latent space through multiplication and addition. They showed that modeling context-specific user-item interactions significantly improve recommendation accuracy.

\subsection{Conditional Recurrent Neural Networks}
RNNs define a function over sequences in some vocabulary. In particular, it is useful to consider a probability density function for the task of sequence modeling. 
Conditional RNN is a class of RNN that assign probabilities to sequences given a representation of conditioning context. 

\subsubsection{Conditioning input representation}
\label{sec:cond-input}
An attempt to include complementary input information into the RNN was proposed in \cite{crnn:mikolov}. The main idea is to condition the hidden and output vectors on a complementary information by concatenating feature vector with input and output vectors. In particular, authors study topic information that comes to complement the input word vector in language modeling setting. Other works proposed including linguistic characteristics such as such as part of speech tags \cite{DBLP:journals/corr/SennrichH16} or syntactic dependency information \cite{DBLP:journals/corr/NadejdeRSDJKB17} about the current word. 

Concatenation assumes no effect of complementary information on input representation. A tighter binding of complementary information into input representation was proposed by Kiros et al. in \cite{mult-context:kiros}. Authors introduced the idea of multiplicative interaction to learn representations of text conditioned on attributes, e.g. author information or language indicator. This resulted in word embedding vectors that capture similarities across different attributes.

\subsubsection{Conditioning hidden dynamics}
\label{sec:cond-dynamics}
Vanilla RNNs limit the number of ways the different inputs can form different state sequences. Therefore, multiple works researched ways of adapting the hidden transitions to the input information. 

In \cite{tensor-rnn:sutskever} the authors propose Tensor RNN that achieves input-dependent dynamics by introducing a transition matrix for each input. It defines a 3-way tensor with separate transition matrix for each input dimension. This solution is impractical due to the size of the resulted tensor. Therefore, the authors propose a low-rank approximation of this tensor into a product of three matrices, called Multiplicative RNN (mRNN). Authors note that such parametrization makes gradient descent learning difficult.

In \cite{mirnn}, authors addressed the issue of input-dependent transitions by proposing a gating type structure, in which the two terms in vanilla RNN formulation are the gates of each other. Specifically, the state-to-state computation is being dynamically rescaled by input-to-state computation. RNN with this structure is referred to as Multiplicative Integration (MI-RNN). Compared to mRNN, MI-RNN offers easier optimization and simpler parametrization.

A recent paper on HyperNetworks \cite{hypernetworks} parametrizes the transition matrix through an auxiliary smaller RNN. In both mRNN and MI-RNN approaches, the weight augmentation terms are produced by a linear operation, while in the HyperNetworks approach, the weight scaling vectors are dynamically produced by another RNN with its own hidden states and non-linearities.

Some of recently proposed RNN architectures use recurrent depth \cite{DBLP:journals/corr/ZhangWCLMSB16}, which is a length of path between recurrent steps. Recurrent depth allows more non-linearity in the combination of inputs and previous hidden states from every time step, which in turn allows for more flexible input-dependent transitions.

\subsection{This work}
As discussed in Section \ref{sec:rnn-for-reco}, RNNs successfully capture the order information in a sequence of user-item interactions. Previous works in CAR presented in Section \ref{sec:car} showed that additional contextual information improves the performance of recommender systems. In this paper, we aim to leverage contextual information in the RNN to further improve sequence modeling. Specifically, we present to two ways the context information could be included into the RNN: (1) by conditioning item representation on the context, see related work in Section \ref{sec:cond-input}, and (2) by conditioning hidden dynamics of the RNN, see related work in Section \ref{sec:cond-dynamics}.


\section{Proposed Approach}
\label{sec:proposedapproach}
In this paper, we are interested in modeling both context-dependent item representation as well as context-dependent dynamics of the RNN. 
By modifying representation of item with context, we aim to capture context-dependent item similarities. Making RNN transitions context-dependent allows to model the changes in the user behavior based on context.
Experimentally we show improved modeling of sequences of user-item interactions when conditioning on type of the interaction and time dimension.

\subsection{Model and optimization}
\label{sec:proposedapproach:model}

\subsubsection{Setup}
We are given sequence of paired inputs $X=\left\{(x_t, c_t)\right\}, t=1 \ldots T$, where $x_t \in R^{V_{x}}$ and $c_t \in R^{V_{c}}$ are one-hot encoded item id and context vectors respectively at time step $t$. 
Our objective is to produce likely continuations of sequences and use them as a recommendation. To this end, we define a probability distribution over sequences $p(X)$.

The joint probability $p(X)$ can be decomposed using the chain rule into a product of conditional probabilities:
\begin{eqnarray*}
p(X) = p(x_1, .., x_T, c_1, .., c_T) = \\ 
\prod_{t=1}^T p(x_t| c_t, x_{<t}, c_{<t}).
\end{eqnarray*}
Thus, our task reduces to the one of modeling the probability of the next item $x_t$ given the current context $c_t$ and the history of items and contexts $\left\{(x_{<t}, c_{<t})\right\}$. 
In this paper, we propose to model $p(x_t| c_t, x_{<t}, c_{<t})$ using RNN.

\subsubsection{Recurrent architecture}
\label{sec:recurrent-architecture}
\begin{figure}[t]
\begin{center}
  \includegraphics[scale=0.5]{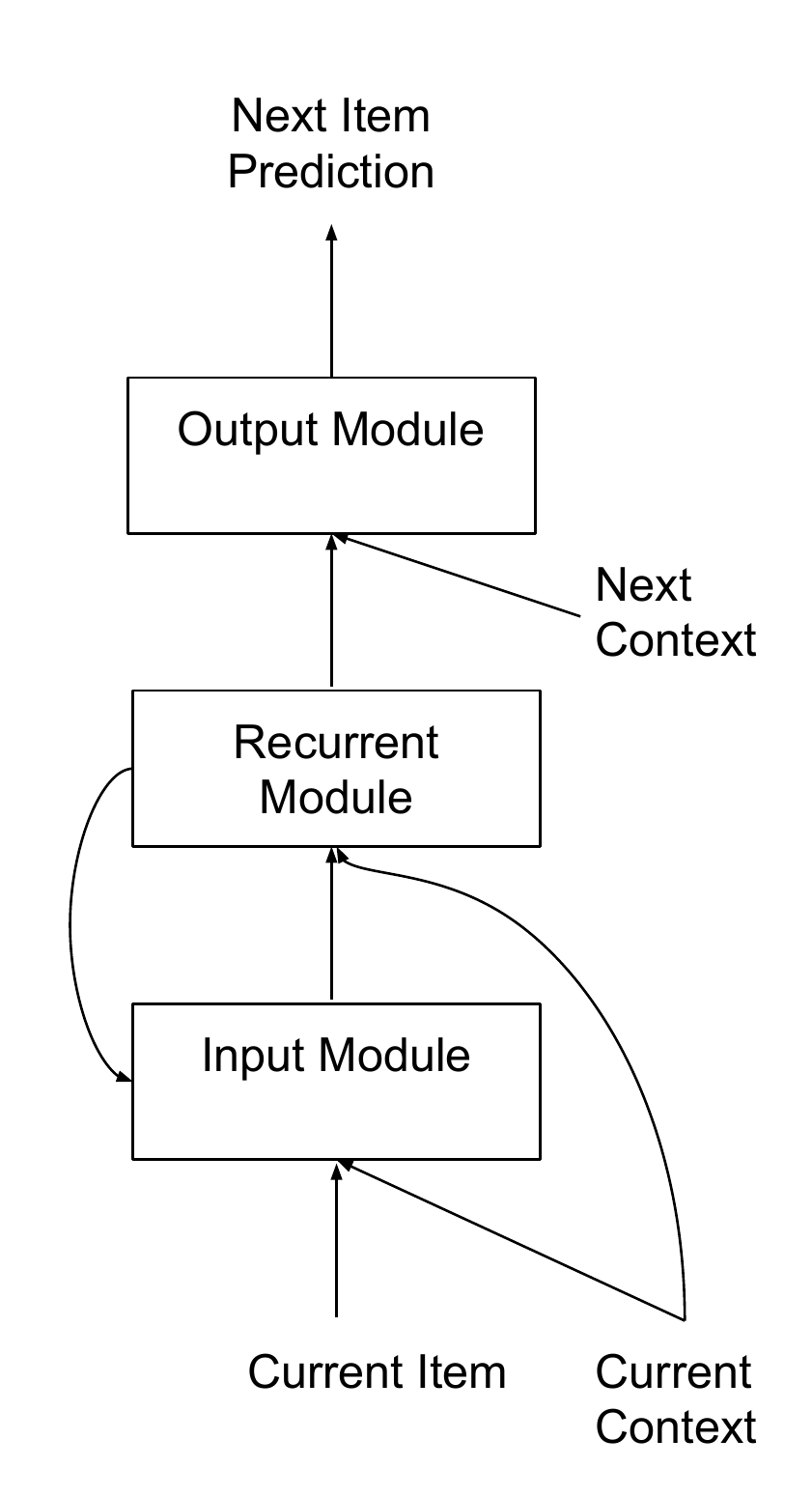}
  \caption{Recurrent architecture for contextual sequence modeling.}
  \label{fig:crnn}  
\end{center}
\end{figure}

We consider the following recurrent architecture, illustrated in Figure \ref{fig:crnn}. At each time step, it consists of input, recurrent and output modules. 

\paragraph{Input module}
Input module is to create the dense input embedding from sparse original input data:
\begin{eqnarray*}
x_t^{embed} = f^{in}(x_t), \\
x_t^c = \theta^{in}(x_t^{embed}, c_t),
\end{eqnarray*}
where $f^{in}$ is item representation function that returns dense embedding vector $x_t^{embed} \in R^{N_x}$ of input item id $x_t$, $\theta^{in}$ is the input integration function that brings context information into the item representation. Then, the final combined representation $x_t^c$ is passed to the recurrent update module.

\paragraph{Recurrent module}
This module updates the hidden state vector with the current input and previous state:
\begin{eqnarray*}
h_t=\phi(x_t^c, h_{t-1})
\end{eqnarray*}
where $\phi$, a is cell module, such as Gated Recurrent Unit (GRU) \cite{gru:cho} or Long Short Term Memory (LSTM) \cite{lstm:schmidhuber}, $h_t \in R^{k}$ is a state vector and $k$ is the number of hidden state dimensions.\\

\paragraph{Output module}
Finally, output module returns a probability distribution over items based on the updated state vector and next context vector:
\begin{eqnarray*}
h_t^c = \theta^{out}(h_t, c_{t+1}), \\ 
o_t=f^{out}(h_t^c),
\end{eqnarray*}
where $\theta^{out}$ is the output integration function that brings the next context vector into the prediction, $f^{out}$ is the output function that returns a vector $o_t \in R^{V_x}$ in original item id space.

The final prediction is the probability distribution over items given by softmax function applied to the output vector $o_t$: 
\begin{eqnarray*}
p(x_t| c_t, x_{<t}, c_{<t}) = softmax(o_t) = \\
\left[\frac{\exp(o_{t, 1})}{\sum_{i \in V_x} \exp(o_{t, i})}, \ldots, \frac{\exp(o_{t, V_x})}{\sum_{i \in V_x} \exp(o_{t, i})}\right],
\end{eqnarray*}
where $o_{t, i}$ is the $i$-th element $o_t$ vector.

In the simplest form, both $f^{in}$ and $f^{out}$ are linear functions. In addition, the embedding matrix is the same on both input and output. This experimentally provided better results and supports findings in \cite{tying-word-vectors:inan}: 
\begin{eqnarray*}
x_t^{embed} = Vx_t, \\
o_t = Vh_t^c.
\end{eqnarray*}

\subsubsection{Conditioning on context}

\paragraph{Context-dependent input/output representation}
\label{sec:context-input}
We explore three input integration functions $\theta^{in}$:
\begin{itemize}
	\item concatenation $\theta^{in}(x_t) = [x_t; c_t]$,
	\item multiplicative interaction $\theta^{in}(x_t) = x_t \odot C c_t$,
	\item concatenation and multiplicative interaction $\theta^{in}(x_t) = [x_t \odot C c_t; c_t]$,
\end{itemize}
where ``$\odot$'' denotes element-wise multiplication. For output integration function $\theta^{out}$, we consider the same choices of functions but applied to state vector $h_t$ instead of input vector $x_t$.

Concatenation assumes no effect of context on item representation and is similar to \cite{crnn:mikolov}.

Multiplicative interaction offers tighter binding of context into item representation and is similar to \cite{mult-context:kiros}.

We note that the resulted integrated input representation implicitly effects the recurrent update. In the next section, we explicitly parametrize the recurrent update with the context.

\paragraph{Context-dependent hidden dynamics}
\label{sec:context-dynamics}
Most of popular cell structures \cite{gru:cho}, \cite{lstm:schmidhuber} share the same computational block:
\begin{eqnarray*}
g(W[x_t^c; h_{t-1}] + b),
\end{eqnarray*}
where $g$ is activation function, such as sigmoid or hyperbolic tangent, $W \in R^{N_x^c \times k}$ is a transition matrix, $b \in R^k$ is a bias vector. We modify this block to introduce context-dependent transitions:
\begin{eqnarray*}
g(W^c[x_t^c; h_{t-1}] + b).
\end{eqnarray*}
Having a separate transition matrix for each context is impractical due to its size. Therefore, we introduce the multiplicative term to condition the transition matrix on context: 
\begin{eqnarray*}
g(W[x_t^c; h_{t-1}] \odot U c_t + b).
\end{eqnarray*}
We refer to this structure as \textit{Context Wrapper}.

In experiments we implemented Context Wrapper for GRU cell:
\begin{eqnarray}
\label{eq:context-wrapper-1}
u_t &=& \sigma(W_u[x_t^c; h_{t-1}] \odot U_u c_t + b_u), \\
r_t &=& \sigma(W_r[x_t^c; h_{t-1}] \odot U_r c_t + b_r), \\
\hat{h_t} &=& \tanh(W_h[x_t^c; h_{t-1} \odot r_t] \odot U_h c_t + b_h), \\
h_t &=& (1-u_t)\odot h_{t-1} + u_t \odot \hat{h_t}, 
\label{eq:context-wrapper-4}
\end{eqnarray}
where $\sigma$ denotes the sigmoid function, $\tanh$ is the hyperbolic tangent, $u_t$ and $r_t$ are update and reset gates respectively.

\subsubsection{Optimization objective}
We train by minimizing the negative log-likelihood of the training data:
\begin{eqnarray}
\label{eq:loss}
L = - \sum_{i=1}^N \sum_{t=1}^{T^{(i)}} \log p(x^{(i)}_t| c^{(i)}_t, x^{(i)}_{<t}, c^{(i)}_{<t}) = - \sum_{i=1}^N \sum_{t=1}^{T^{(i)}} \log o^{(i)}_{t, x^{(i)}_t},
\end{eqnarray}
where $N$ is the number of training sequences, $T^{(i)}$ is the length of $i$-th sequence, $o_{t, x_t}$ is element of $o_t$ that corresponds to item id $x_t$.

\section{Experiments}
\label{sec:experiments}
The experimental section is organized as follows. First, we describe the evaluation task, success metrics, the baselines and the configurations of contextual models. Then, we present datasets for experiments. Finally, we report the results of the experiments.

\subsection{Setup}
We evaluate the recommendation methods on the next event prediction task. We focus on evaluating the quality of top K recommendations. As evaluation metric, we use Recall at K (Recall@K) averaged over all events. Recall at K is the proportion of time the test item appears in the top K list of predicted items. This metric was reported to be correlated with online metric, namely click-through rate \cite{car:balazs}.

\subsection{Baselines}
We compare against following baselines that we define using their recurrent update module:
\begin{itemize}
\item \textit{CoVisit}: $h_t = x_t^{embed}$, only takes into account the last visited product, does not take into account earlier history
\item \textit{BagOfItems}: $h_t = h_{t-1} + x_t^{embed}$, computes sum of embedding vectors of historical products, does not take into account the order
\item \textit{RNN without context}: $h_t = \phi(x_t^{embed}, h_{t-1})$, models the order of historical products but not associated context
\item \textit{RNN on ItemId x EventType}: $h_t = \phi(x_t \times c_t, h_{t-1})$, models context through a item vector per each event type, ``$\times$'' denotes the Cartesian product.
\end{itemize}
For all experiments, the cell function $\phi$ is the GRU cell.

\subsection{Contextual RNNs}
Based on the recurrent architecture described in Section \ref{sec:recurrent-architecture}, we experiment with the following configurations:
\begin{itemize}
	\item \textit{Mult-GRU-RNN}: RNN with GRU cell, multiplication on input and output modules, presented in Section \ref{sec:context-input}
	\item \textit{Concat-GRU-RNN}: RNN with GRU cell, concatenation on input and output modules, presented in Section \ref{sec:context-input}
	\item \textit{Concat-Mult-GRU-RNN}: RNN with GRU cell, concatenation and multiplication on input and output modules
	\item \textit{Concat-Mult-Context-RNN}: same as above, but with Context Wrapper for GRU cell, described in Section \ref{sec:context-dynamics}, Equations \ref{eq:context-wrapper-1}-\ref{eq:context-wrapper-4}.
\end{itemize}

\subsubsection{Hyper-parameters}
We fixed the size of item embedding vector and the size of RNN hidden state to 100. For optimization of the loss function in Equation \ref{eq:loss}, we use the Adam algorithm with squared root decay of learning rate from 0.01 to 0.001. For all models, the batch size was set 256 and number of training iterations to 10,000.

\subsection{Datasets}
We experimented on two datasets. 

The first dataset is publicly available YooChoose dataset \cite{dataset:yoochoose} introduced for RecSys challenge 2015. It is a collection of sessions of user clicks and purchases on multiple e-commerce websites over a period of 6 months from April 2014 till September 2014. Each user session form a separate sequence and users are not identifiable between sessions.

The second dataset is proprietary dataset that contains user browsing and purchasing activity on multiple e-commerce websites from diverse verticals over the period of 3 months from December 2016 till March 2017. We refer to this dataset as Internal dataset. All activity of a user on one e-commerce website forms a sequence. For this dataset, we replaced product ids with less that 5 occurrences with a single id.

The statistics of the datasets are presented in Table~\ref{tab:dataset-stats}. The Internal dataset has much heavier long tail in terms of number of distinct items and contains less sessions than YooChoose dataset.

We filter out sequences of length 1 and keep 20 latest events in each sequence. For YooChoose dataset, we split sequences into the consequent training, validation and test periods. The duration of validation and test periods are of 1 week each. For Internal dataset, we held out 20\% of randomly selected users for each of validation and test sets, accounting in total for 40\% of hold-out users. 

\begin{table}
\centering
\begin{tabular}{lrr}
\toprule
\textbf{Dataset} & \textbf{YooChoose} & \textbf{Internal}\\
\midrule
Training sessions & 8M & 2.3M\\
Validation sessions & 295K & 760K\\
Test sessions & 173K & 700K\\
Number of distinct items & 37K & 370K\\
\bottomrule
\end{tabular}
\caption{Datasets statistics.}
\label{tab:dataset-stats}
\end{table}

\subsubsection{Contextual features} 
We consider the following contextual features in our experiments:
\begin{itemize}
	\item Time: timestamp in seconds broken into discrete features -- month, hour and day of week,
	\item Time difference: time since last event in seconds discretized at $\log_2$ scale and with maximum value of 20,
	\item Event type: in the case of YooChoose dataset, event type is either product view or sale; in the case of Internal dataset, event type also contains ``add to basket'' event.
\end{itemize}
These discrete contextual features were represented as one-hot encoded vectors and concatenated.

\subsection{Results}
Table \ref{tab:dataset-results} summarize the global performance of CRNNs versus baselines in terms of Recall@10. All CRNN models improve over the best baseline (results statistically significant at 95\% confidence level). The biggest overall improvement across CRNNs is given by the Concat-Mult-Context-RNN model that internalizes context into the input representation and the hidden dynamics of the RNN. On input/output representation, concatenation and multiplication together bring more uplift than separately. Among two, multiplication alone has more impact on performance on YooChoose dataset than concatenation. On Internal dataset, their performance is similar.

The best performing baseline is GRU RNN without context that outperforms the CoVisit baseline that only considers the last product and the BagOfItems baseline that does not take into account the order of items. \footnote{We were not able to train the RNN on ItemId x EventType baseline as training diverged on both datasets. The reason for this was sparsity introduced by the Cartesian product that lead to the numerical instability of the loss function.}

\begin{table*}
\centering
\begin{tabular}{lrr} 
\toprule
\textbf{Model} & \textbf{YooChoose} & \textbf{Internal dataset}\\
\toprule
\textit{Baselines}\\
CoVisit & 0.374 & 0.329\\
BagOfItems & 0.443 & 0.354\\
GRU RNN without context & \textbf{0.562} & \textbf{0.454}\\
\midrule
\textit{Contextual RNNs}\\
Mult-GRU-RNN & 0.589 (+4.8\%) & 0.467 (+3.0\%)\\
Concat-GRU-RNN & 0.579 (+2.9\%) & 0.468 (+3.0\%)\\
Concat-Mult-GRU-RNN & 0.592 (+5.3\%) & 0.472 (+3.9\%)\\
Concat-Mult-Context-RNN & \textbf{0.600 (+6.6\%)} & \textbf{0.474 (+4.3\%)}\\
\bottomrule
\end{tabular}
\caption{Comparison of performance in terms of Recall@10 of CRNNs versus non-contextual models. Uplift in percents over the best baseline is shown in parentheses. All results are statistically significant based on 95\% confidence interval obtained from 30 bootstraps.}
\label{tab:dataset-results}
\end{table*}

\subsubsection{Analysis}
We analyze the performance gain of contextual models by looking at two projections of interest:
\begin{itemize}
	\item New or historical item. As a recommendation system, we are interested in recommending items that user has not seen previously (referred to as ``new'') as opposite to historical items. We believe that recommending relevant previously unseen items shows the model ability to understand the user sequence.  
	\item Event type of the next item. As a business, we are also interested in predicting items that will be eventually bought by the user.
\end{itemize}

Table \ref{tab:evt-projection} and \ref{tab:is-hist-projection} demonstrate that CRNNs succeed by large margin in both recommending more relevant non-historical products and correctly predicting the sold item than non-contextual models. 

Specifically, from Table \ref{tab:evt-projection} we observe significant improvement in Recall@10 on sale events than on view events on both YooChoose and Internal datasets. The biggest improvement of $+12\%$ on YooChoose dataset and $+6\%$ on Internal dataset comes from the Concat-Mult-Context-RNN model. We note that sale events represent less than 5\% of all events on both datasets. Thus, we conclude that more advanced contextual models improve on rare events that are hard to capture with simpler models. 

Similar analysis in Table \ref{tab:is-hist-projection} shows that CRNNs recommend up to $+10\%$ more relevant non-historical items than models without contextual information. The biggest performance improvement is again given by the Concat-Mult-Context-RNN model. We note that the baseline fit well on historical items but it is less performant on non-historical ones. We conclude that CRNNs improve on harder non-historical examples.

\begin{table*}
\centering
\begin{tabular}{lrrrrr}
\toprule
\textbf{Model} & \multicolumn{2}{c}{\textbf{YooChoose}} & \multicolumn{3}{c}{\textbf{Internal dataset}}\\
& View & Sale & View & Sale & Basket \\
\midrule
Baseline Recall@10 & 0.560 & 0.602 & 0.355 & 0.729 & 0.683 \\
\midrule
Mult-GRU-RNN & 4.5\% & 8.9\% & 2.4\% & 3.1\% & 4.0\% \\
Concat-GRU-RNN & 2.7\% & 6.1\% & 2.9\% & 3.2\% & 3.3\%  \\
Concat-Mult-GRU-RNN & 4.9\% & 11.0\% & 3.4\% & 4.2\% & 4.9\% \\
Concat-Mult-Context-RNN & \textbf{6.2\%} & \textbf{12.5\%} & \textbf{4.0\%} & \textbf{6.1\%} & \textbf{4.6\%}  \\
\bottomrule
\end{tabular}
\caption{Uplift in Recall@10 of CRNNs over the best baseline broken down by event type of the next item. All results are statistically significant based on 95\% confidence interval obtained from 30 bootstraps.}
\label{tab:evt-projection}
\end{table*}

\begin{table*}
\centering
\begin{tabular}{lrrrrr}
\toprule
\textbf{Model} & \multicolumn{2}{c}{\textbf{YooChoose}} & \multicolumn{3}{c}{\textbf{Internal dataset}} \\
& New & Historical & New & Historical \\
\midrule
Baseline Recall@10 & 0.422 & 0.868 & 0.139 & 0.802 \\
\midrule
Mult-GRU-RNN & 7.7\% & 1.5\% & 6.4\% & 2.4\% \\
Concat-GRU-RNN & 4.1\% & 1.5\% & 8.0\% & 2.2\% \\
Concat-Mult-GRU-RNN & 8.0\% & 2.1\% & 8.2\% & 3.3\% \\
Concat-Mult-Context-RNN & \textbf{10.2\%} & \textbf{2.4\%} & \textbf{9.4\%} & \textbf{3.3\%} \\
\bottomrule
\end{tabular}
\caption{Uplift in Recall@10 of CRNNs over the best baseline broken down by whether the next item has been previously seen by the user (historical) or not (new). All results are statistically significant based on 95\% confidence interval obtained from 30 bootstraps.}
\label{tab:is-hist-projection}
\end{table*}

\begin{figure*}
\begin{center}
  \includegraphics[scale=0.8]{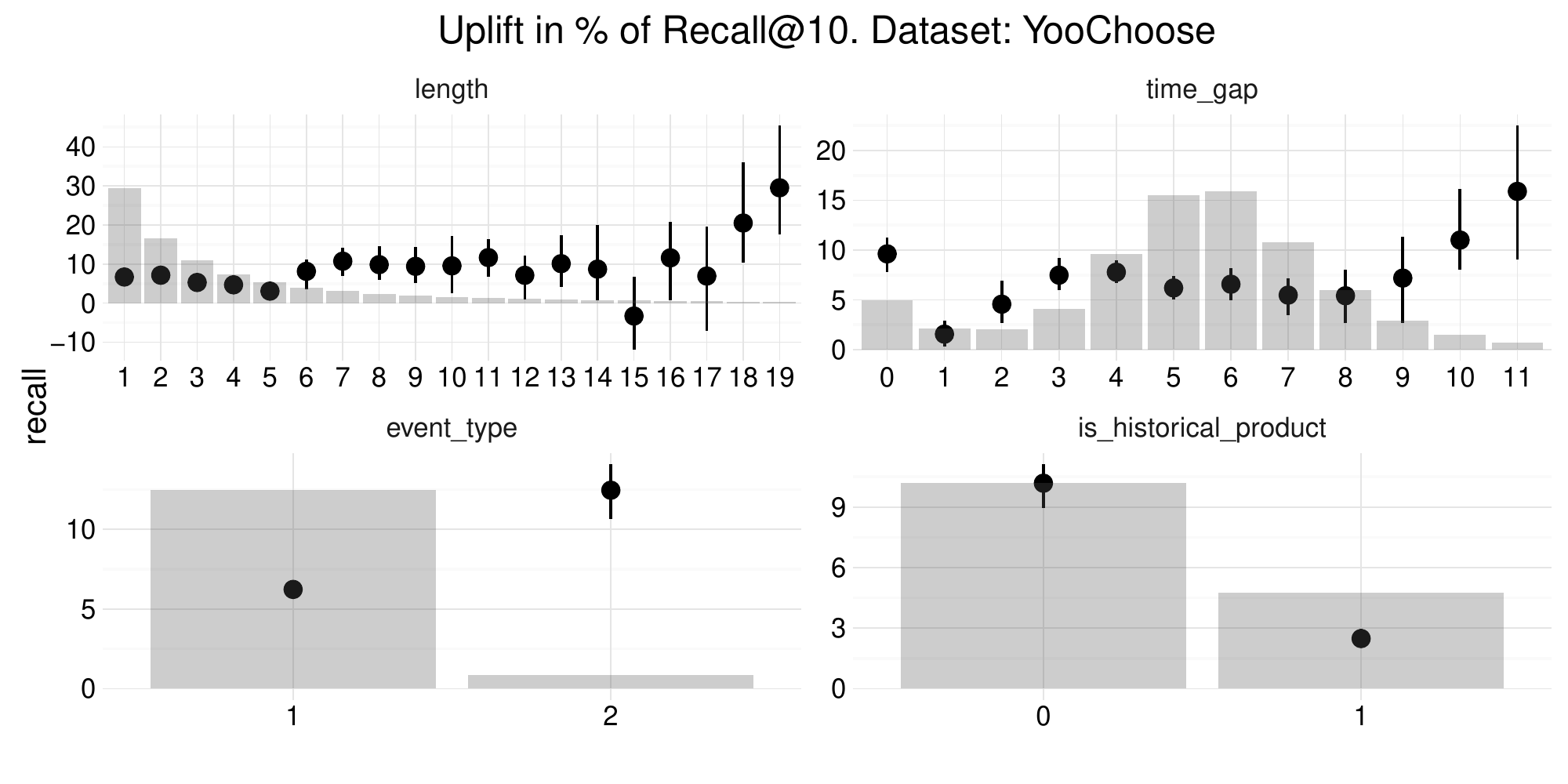}
	\includegraphics[scale=0.8]{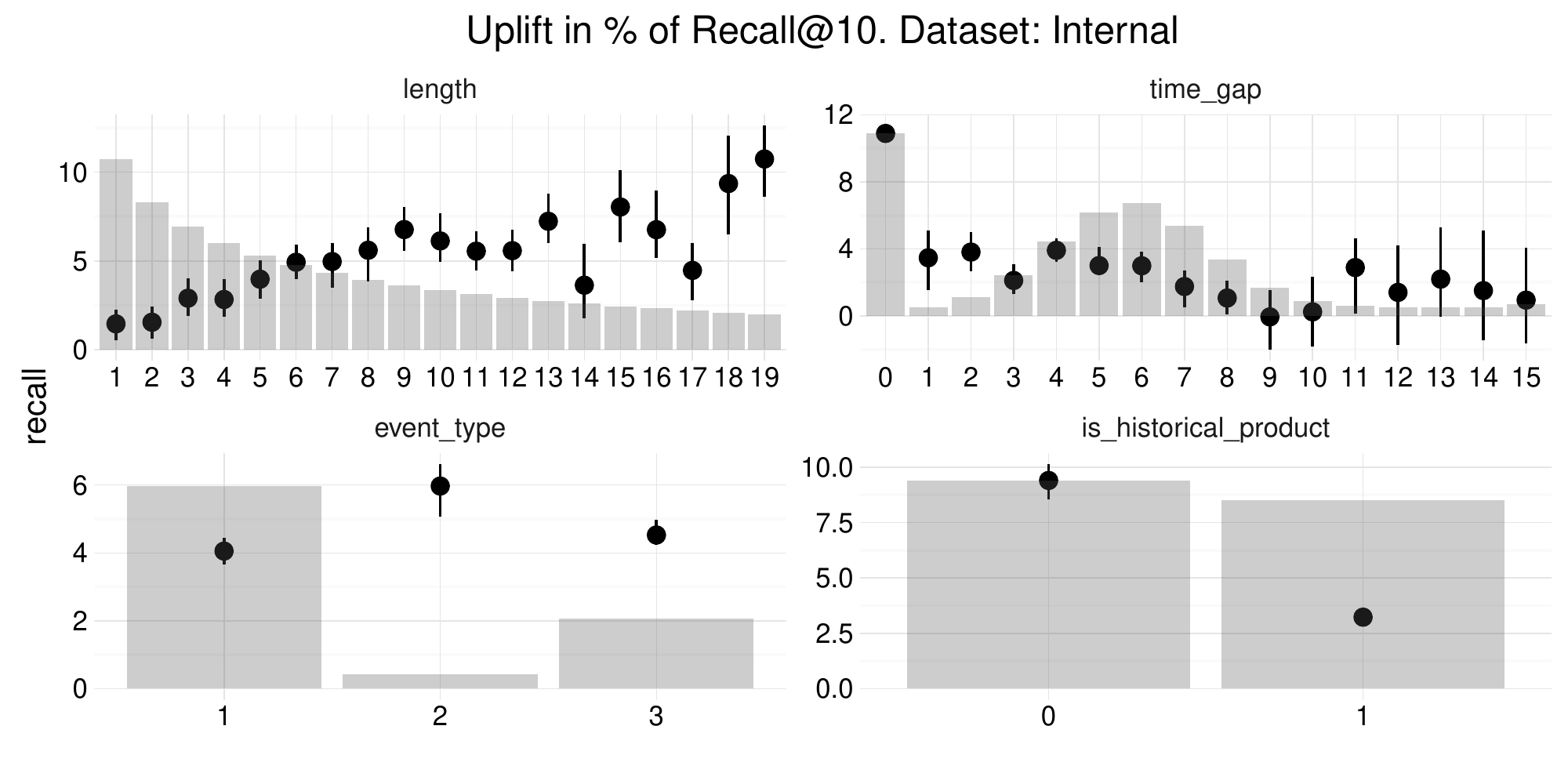}
  \caption{Uplift in percents of Recall@10 of the best CRNN over the best baseline as a function of the event type of next item, new versus historical item, time since the last event and sequence length. Event type legend: 1 - view, 2 - sale, 3 - basket. Errors bars are based on 95\% confidence interval obtained from 30 bootstraps. Gray bars indicate the relative volume of test examples under a particular value of projection.}
  \label{fig:best-contextrnn-projections}
 \end{center}
\end{figure*}

We visualize these findings in Figure \ref{fig:best-contextrnn-projections}. In addition, we show the break-down of uplift in Recall@10 over time since the last event and sequence length. As can be seen from the figure, the longer the sequence, the bigger gets the performance improvement. In terms of time difference between consecutive events, the model improves the most on very recent events (time gap is less than 2 seconds).

\section{Conclusions}
\label{sec:conclusion}

In this paper we introduced the problem of \emph{Contextual Sequential Modeling for Next Item Recommendation} and proposed as a solution the new family of \emph{Contextual Recurrent Neural Networks (CRNN)} architectures. The proposed solution achieves significantly better results against sequential and non-sequential state-of-the-art baselines. The improvements are especially large on hard cases of next item prediction, such as sale events and non-historical items.

As future work, we intend to extend this study to prediction of next K items in the sequence, as opposite to just the next one. Predicting the future context together with the future item id is another direction we plan to research. 

\bibliographystyle{ACM-Reference-Format}
\bibliography{literature}

\end{document}